# Quantum phase transitions in magnetism and superconductivity: emergent spin topology seen with neutrons

W.J.L. Buyers,[a,b*] C. Stock,[c] Z. Yamani,[a] R.J. Birgeneau,[b,c] R. Liang,[b,d] D. Bonn,[b,d] W.N. Hardy,[b,d] C. Broholm,[c] R.A. Cowley,[f] and R. Coldea[f]

[a]*Canadian Neutron Beam Centre, National Research Council, Chalk River, ON, K0J 1J0, Canada*

[b]*Canadian Institute for Advanced Research, Toronto, Ontario, Canada M5G 1Z8*

[c]*Johns Hopkins University, Baltimore, MD USA 21218*, [d]*University of California at Berkeley, Berkeley, CA 94720*

[e]*University of British Columbia, Vancouver, B. C., Canada V6T 2E7*, [f]*Clarendon Laboratory, Oxford, United Kingdom OX1 3PU*

**Abstract**

Magnetic spins and charges interact strongly in high-temperature superconductors. New physics emerges as layers of copper oxide are tuned towards the boundary of the superconducting phase. As the pseudogap increases the characteristic spin excitation energy decreases. We show that our well-annealed $YBa_2Cu_3O_{6+x}$ ($YBCO_{6+x}$) single crystals are orthorhombic and superconducting but not antiferromagnetically ordered. Near the critical concentration for superconductivity for x = 0.35 the spins fluctuate on two energy scales, one a relaxational spin response at ~2 meV and the other a slow central mode that is resolution-limited in energy (<0.08 meV) but broad in momentum. The gradual formation on cooling of a central mode over a range of momenta suggests that the spin ground state from which coherent superconducting pairing emerges may be quantum disordered. We show that $YBCO_{6.35}$ adopts a homogeneous state that consists of highly-organized frozen sub-critical three-dimensional spin correlations. The continuous spin evolution indicates that a single quantum state occurs in contrast to claims from site-based probes that lightly doped YBCO undergoes a transition to antiferromagnetic Bragg order followed by a sharp transition to a cluster glass phase. For x=0.35, where $T_c$ = 18 K is reduced to 1/5 of $T_{cmax}$, the spin ground state is reached without a sharp transition and consists of short correlations extending over only 8 Å between cells and 42 Å within the planes. Polarized neutrons show the angular spin distribution to be isotropic unlike the AF insulator. Since moment is conserved we interpret this as evidence for hole-induced spin rotations rather than decay.

---

[*] Corresponding author. Tel.: +1-613-584-3311; fax: +1-613-584-4040; e-mail: william.buyers@nrc.gc.ca

## 1. Introduction

As carriers are doped into cuprate layers the phase changes from an antiferromagnetic (AF) Mott insulator to a d-wave superconductor when the hole doping reaches p~0.055. For $La_{2-x}Sr_xCuO_4$ (LSCO), the superconducting (SC) phase emerges from an insulating precursor phase with frozen short-range static magnetic order separating the AF from the SC phases [1,2]. In contrast, evidence exists from thermal conductivity that the $YBCO_{6+x}$ precursor phase is metallic [3], an idea challenged by the log(1/T) resistive behaviour [4]. However the latter samples in [4] are more disordered; for x=0.35 they give $T_c = 0$ K whereas the well-annealed crystals we have studied are superconducting below $T_c = 18$ K for the same oxygen doping. Thus the precursor state to superconductivity may be different in YBCO from LSCO and the spin structure may be reorganized to accommodate metallic carriers while avoiding the localization associated with long-range AF order.

The singlet formation energy for screening of a $Cu^{2+}$ spin by a hole on the oxygen ligands in the precursor phase is believed to be extremely large [5]. Whether these singlet hole states are locally paired in the precursor state remains an issue, but there is no doubt that the doping by holes reorganizes the spins and their orientations in a major way.

## 2. Incommensurate modulation, resonance, velocity

Experiments at Chalk River [6] on 70% detwinned $YBCO_{6.5}$ in the ortho-II structure, where every second chain is oxygen filled along the b-axis, show that for energies up to 24 meV the incommensurate dynamic modulation occurs along a* but not b* (Fig.1). We find that 1D incommensurate peaks along a* fully account for observation taking account of the detwinning ratio in agreement Ref. [7]. The suggestion by Hinkov *et al* [8] that the incommensurate modulations lie on a circle, with intensity preferentially peaked along a* is based on data at energies that are a large fraction of the resonance energy (83% for $YBCO_{6.6}$). Since the hour-glass shape of the dispersion [9] converges to the commensurate resonance momentum $(\pi, \pi)$, it may not be surprising that the pattern just below resonance approaches a circle. The data of Stock *et. al.* [9] displays the limiting low-energy behaviour at 12, 8 and 6 meV, the last being only 20% of the resonance energy, low enough that the incommensurate modulation along a* can fully develop. For the antiphase domain dynamic stripe model, $YBCO_{6.5}$ gives charge stripes of length 6 cells, with a spin repeat distance of 12 cells.

At higher energies the incommensurate peaks move to the two-dimensional (2D) commensurate AF momentum $\mathbf{Q}=(\pi, \pi)$. There the spin response is gapless in the normal phase and linear in ω, $\chi''(\mathbf{Q},\omega)/\omega = 0.014$ $\mu_B^2/meV^2$ at low energies (Fig 2 lower panel).

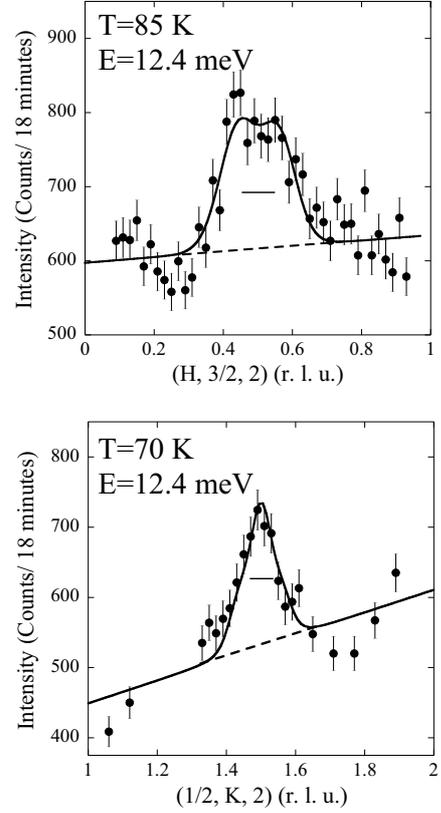

Figure 1 Incommensurate modulation along a* (upper panel) not b* (lower panel) in $YBCO_{6.5}$

Surprisingly the spin susceptibility at **Q** in the normal phase grows inexorably on cooling [6], unlike the declining NMR relaxation rate $1/(T_1T)$ as shown in Fig. 3. It might have been thought that the effectiveness of the SC order to produce a sharp suppression below $T_c$, as seen in the 12 meV (3 THz) neutron spin susceptibility, would be even greater at the extremely low <100 MHz frequency of the NMR experiments, but such is not the case. It is interesting that the NMR relaxation rate begins to decline below the temperature where the incommensurate modulations, and concomitantly the peak at the resonance energy, first appear in the neutron scattering. We believe that, as the spin spectrum hardens with the appearance of a resonant-like feature in the normal phase, so the low-frequency fluctuations that cause NMR relaxation decline.

In the superconducting phase the response is suppressed at low energies and transferred to the resonant feature peaked at 33 meV (Fig. 2). We observed a strongly asymmetric spectrum rising slowly to a resolution-limited cutoff above the peak. The spectral form is reminiscent of the density of SC d-symmetry pair states, with few states near the nodes and many near the maximum gap. A vestige of a broadened resonant feature persists to the same temperature $2T_c \sim 120$ K as the incommensurate modulations [6], as if local SC pairs exist in the normal phase. On cooling we are witnessing, with conservation of moment [6], the transfer of spin response from low to high frequencies as incoherent pairs and the pseudogap form below $T^* \sim 120$ K. The reduced charge density then allows incommensurate modulations to develop at low energy.

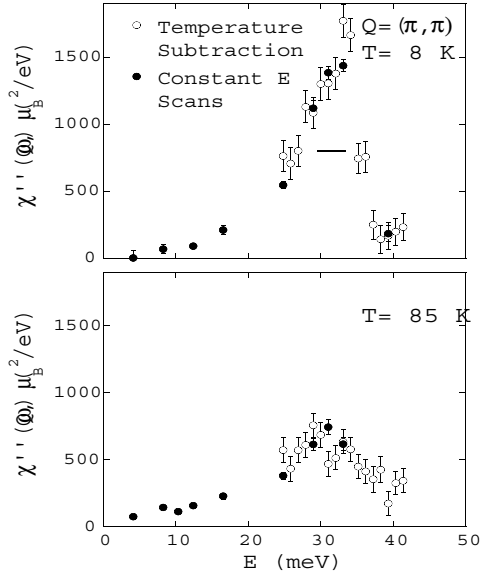

Figure 2 Spin susceptibility of YBCO6.5 showing the d-like resonance in the superconducting phase (upper panel) and the residual resonance in the gapless normal phase (lower panel).

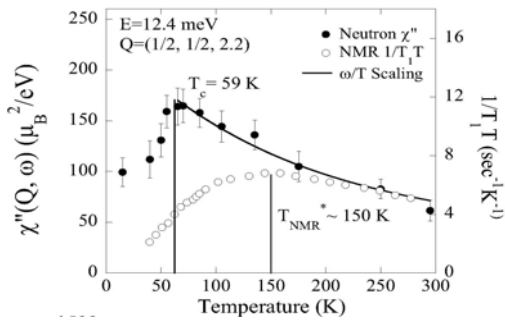

Figure 3 Comparison of spin susceptibility at $\mathbf{Q} = (\pi,\pi)$ measured directly with neutron scattering in upper panel (filled circles) with the NMR copper spin relaxation rate (open cicles). There is no evidence for opening of a pseudogap below $T^*$ in the spin susceptibility at 12.4 meV.

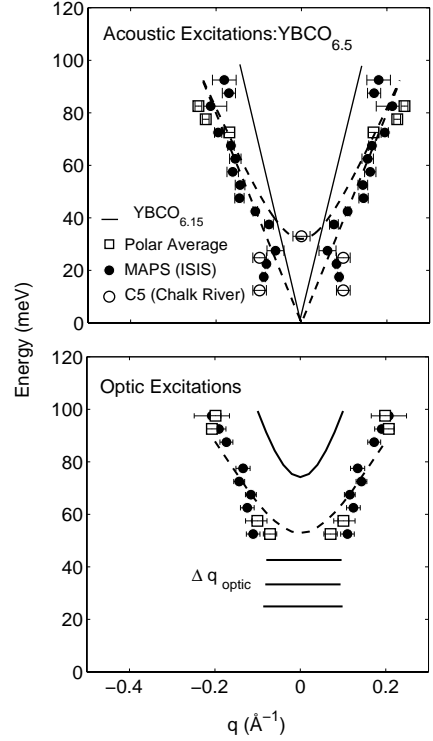

Figure 4 Dispersion of peaks in the spin response up to 100 meV. Below the 33 meV resonance the incommensurate modulations give an umbrella-shaped dispersion. The high energy acoustic paramagnon velocity (upper panel) is reduced to ~60% (broken line) of that of the AF insulator (solid line).

The high-energy scattering measured at MAPS at ISIS provides an overview (Fig. 4) of the hour-glass shaped dispersion of peaks in the spin response [9]. The maxima in $\mathbf{Q}$ for constant-E arise from paramagnons and are not resonant in energy. An unexpected result is that the optic weight, whose onset occurs at ~70 meV for the AF insulator, now can be seen as low as 25 meV, below the acoustic resonance. This result has been confirmed with triple axis techniques at NRU reactor in Chalk River [9].

The pattern in momentum in the a-b plane of the high-energy excitations is isotropic for $YBCO_{6.5}$. A constant wave-vector radius is observed in all a-b directions up to 100 meV. It is the expected form for a cone of paramagnons crossing a constant-energy slice. A different box-shaped pattern was observed in YBCO$_{6.6}$ [10], a doping where more phases exist with lattice modulations that might influence the spin pattern [11]. We find the paramagnon velocity of 400 meV-Å is reduced to ~ 60% of the AF insulator spin-wave velocity. Nonetheless the total moment up to 120 meV of 0.3 $\mu_B^2$ is comparable to that of the insulator (0.2 $\mu_B^2$) when allowance is made for the reduced exchange constant (0.2/0.6~0.3 $\mu_B^2$). No moment, to an accuracy of the 9% hole doping, is therefore lost by doping or by superconductivity.

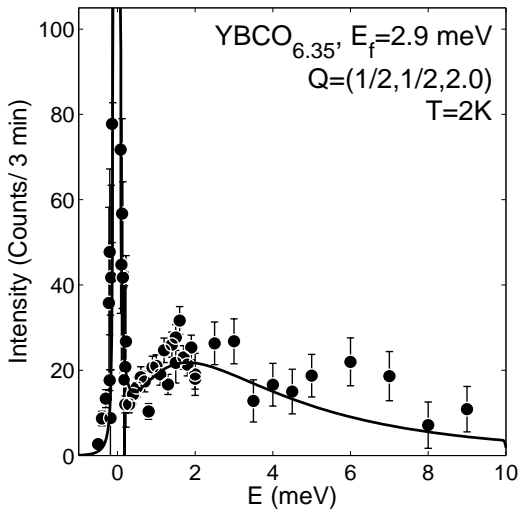

Figure 5 Two energy scales, a narrow central mode with FWHH<0.08 meV, and faster relaxational excitations peaked at ~2 meV.

## 3. Coupled central mode and spin relaxation on approach to the superconducting phase boundary.

There have been substantial improvements over the past decade in crystal quality and particularly in the ability to achieve good oxygen ordering. With the careful attention to growth and annealing at UBC we now know that the superconducting boundary occurs well below the 6.40 doping of the early materials [12] and closer to 6.31. In particular YBCO$_{6.35}$ is found to be superconducting below $T_c$=18 K, to be orthorhombic so as to provide long chains that are a good sink for electrons, and to exhibit no magnetic Bragg peaks. The spin correlations are short-range and undergo a gradual freezing on cooling [13]. There is no evidence for the sharp transitions adduced to interpret site-based µSR data [14, 15].

In YBCO$_{6.35}$ where p~0.06, close to the lower boundary of the superconducting phase, the spins have been shown [13] to fluctuate on two energy scales as shown in Fig. 5. The faster relaxation rate of ~2.5 meV describes overdamped spin excitations and is an order of magnitude lower than the 33 meV resonance in YBCO$_{6.50}$. It may be a vestige of an overdamped resonance and as such would support early predictions of a spin soft mode[16]. The much slower central mode has a linewidth that decreases at low temperatures to less than the resolution limit of 0.08 meV.

Although central modes have been inferred in previous work on LSCO and YBCO, constant-**Q** spectra such as Fig. 5 directly showing the two timescales in a single scan have not been available for a superconductor. For LSCO the sample was non-superconducting [17] and for YBCO the phase was in addition antiferromagnetically ordered with localized spins [12]. The results of Ref. [13] show that the nascent superconducting metallic phase itself is characterized by slow short-ranged spin dynamics.

The faster spin excitations transfer their weight on cooling to the central mode. Thus the two spectral features arise from a single phase of YBCO and not from phase separation. Furthermore, the total moment integral up to 32 meV is conserved and lies close to the integral of the parent AF insulator over the same energy. It shows that holes and superconductivity do little to destroy the total moment, but greatly disturb the spin order.

The central mode is strongest at integer-centred peaks along (1/2, 1/2, L) yet the central mode is not a 3D Bragg peak. The spin correlations are short, 42 ± 5 Å or ~10 cells within the plane, and much shorter, 8 ± 2 Å along the c-direction based on our coupled bilayer model [13]. In contrast to claims from µSR polycrystal experiments [14, 15] that the spins in YBCO undergo phase transitions and show disorder near the superconducting boundary, we find the spin evolution to be smooth and highly correlated. The diffraction pattern shows that the spins in the bilayer sheets are antiferromagnetically coupled, i.e., their phase is $\phi = \pi$. Any sliding of the spin density in one layer relative to the other would be clearly visible. Likewise any phase difference between cells along c away from the in-phase $\phi = 0$, would cause diffraction peaks at non-integer L. Thus the regions of correlated spin density in one plane maintain an antiferromagnetic coupling to the other plane of the bilayer and an in-phase relation at the same (x,y) from cell to cell. This is precisely the form of spatial fluctuation that precedes a critical quantum transition to 3D antiferromagnetism. The correlation range is short and strongly sub-critical along c.

If the spins were disordered in clusters as claimed [15] the regular integer-centred pattern would not be observed. Neither disordered localized clusters, nor a cluster spin glass as inferred from muon studies, are in accord with experiment. The spins clearly gradually organize into a pattern where the susceptibility at 3D AF momenta (½ ½

L=n) is a maximum. The frozen 3D spin pattern we observe shows that a single phase exists in which superconductivity and glassy spins coexist.

The central mode strength grows smoothly on cooling (Fig. 6). There is no evidence for the several 'transitions' inferred to interpret muon data on polycrystals [15], the SC $T_c$, an AF Néel transition $T_N$, and a freezing transition $T_F$. It is possible that the narrowing central mode enters the ~10 MHz muon range rather quickly in T and gave the impression of a transition. It is also true that there is a large internal surface area in polycrystals where strain might induce spin localization in entire grains. We believe our results give the canonical behaviour of a doped $CuO_2$ plane.

The spin correlations are robust near the lower edge of the superconducting phase, for the amplitude of the central mode (Fig. 6 60 K background subtracted ) and of the spin excitations in $YBCO_{6.35}$ show no anomaly upon entry to the SC phase. This contrasts with $YBCO_{6.5}$, whose stronger superconducting order and weaker spin fluctuations without a central mode cause spin suppression (Fig. 3) to occur below $T_c$ [6].

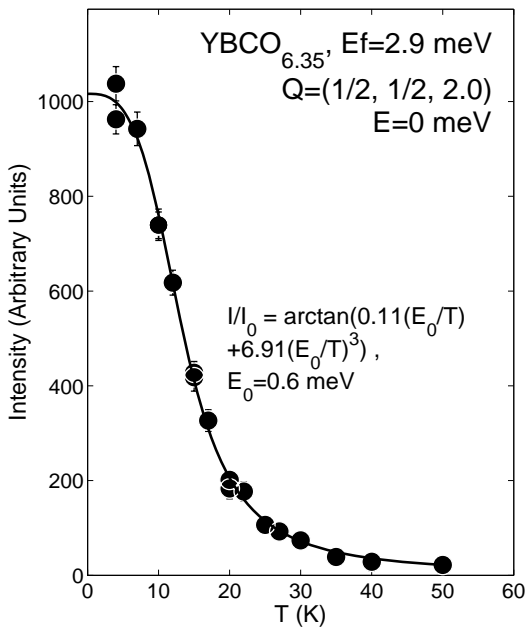

Figure 6 The central mode grows strongly below 30 K but continuously without a transition. The finite-range spin correlations are robust against the onset of superconducting order at 18 K and saturate at $E_0/k_B$=7 K, well below $T_c$ The background at 60 K is subtracted.

The state differs from the insulating antiferromagnet in a surprising way: the spin polarization of the slow central mode and the excitations are isotropic as in a paramagnet [13]. If the system were to lie close to a regular 3D classical temperature AF transition, we would expect the out-of-plane excitations to occur primarily in the zz channel. As we approach the quantum transition this is clearly not the case. In polarized measurements the spin-flip scattering in a horizontal field (HF) and vertical field (VF) exhibit a ratio HF/VF that is consistently equal to 2 [13]. This is as true for the central mode as for the relaxational spin excitations, and it remains so when we vary the ratio of xx to zz by changing L. It would have been easy to detect any xy ordering of the spins for the ratio HF/VF would then rise to ~3 for L=2 and ~6 for L=1, a ratio we would have easily distinguished from 2.

It appears that hole doping produces a much larger effect on spin orientation than it does on the total moment. Spiral and similar forms will do and have been the subject of theoretical speculation [18]. Possibly a spin in a singlet state with a hole forms locally ferromagnetic bonds that lead to spin rotation spreading out from the singlet to a large number of surrounding sites. We know that only a few holes can destroy AF order so many sites per hole must be involved. Spin rotations are enough to give rise to the short correlations observed in our experiments, without reducing the total moment as required by our results.

## 4. Conclusions

As oxygen doping decreases from y=6.50 to 6.35 the 33 meV resonance is replaced by overdamped excitations at an order of magnitude lower energy (~2 meV), concomitant with formation, without passing through a transition, of a central mode at a yet lower order in energy (<0.08 meV). There is no antiferromagnetic Bragg order. The short-range correlations within and between planes reveal a ground state of frozen sub-critical 3D-enhanced spin correlations whose spin directions are isotropic. It appears that the antiferromagnetic fluctuations have frozen in time without orientational order.

The central mode, with resolution-limited energy width and extending over a finite momentum range, breaks the relation of energy to momentum via velocity that is widely used in theory. The hole doping has created a subcritical paramagnetic state that allows superconductivity to occur.